\documentclass{aastex62}

\usepackage{amsmath}
\usepackage{units}
\usepackage{gensymb}

\begin{document}
\title{Tidal-locking-induced stellar rotation dichotomy in the open cluster NGC 2287?}	

\author[0000-0002-3279-0233]{Weijia Sun} 
\affiliation{Kavli Institute for Astronomy \& Astrophysics and
  Department of Astronomy, Peking University, Yi He Yuan Lu 5, Hai
  Dian District, Beijing 100871, China}
\affiliation{Key Laboratory for Optical Astronomy, National
  Astronomical Observatories, Chinese Academy of Sciences, 20A Datun
  Road, Chaoyang District, Beijing 100012, China}
  
\author[0000-0002-3084-5157]{Chengyuan Li}
\affiliation{School of Physics and Astronomy, Sun Yat-sen University, Zhuhai 519082, China}
\affiliation{Key Laboratory for Optical Astronomy, National
  Astronomical Observatories, Chinese Academy of Sciences, 20A Datun
  Road, Chaoyang District, Beijing 100012, China}
  
\author[0000-0001-9073-9914]{Licai Deng}
\affiliation{Key Laboratory for Optical Astronomy, National
  Astronomical Observatories, Chinese Academy of Sciences, 20A Datun
  Road, Chaoyang District, Beijing 100012, China}
\affiliation{School of Astronomy and Space Science, University of the
  Chinese Academy of Sciences, Huairou 101408, China}
\affiliation{Department of Astronomy, China West Normal University,
  Nanchong 637002, China}
  
\author[0000-0002-7203-5996]{Richard de Grijs}
\affiliation{Department of Physics and Astronomy, Macquarie
  University, Balaclava Road, Sydney, NSW 2109, Australia}
\affiliation{Research Centre for Astronomy, Astrophysics and
  Astrophotonics, Macquarie University, Balaclava Road, Sydney, NSW
  2109, Australia}
\affiliation{International Space Science Institute--Beijing, 1
  Nanertiao, Hai Dian District, Beijing 100190, China}

\begin{abstract}
Stars spend most of their lifetimes on the `main sequence' (MS) in the
Hertzsprung--Russell diagram. The obvious double MSs seen in the
equivalent color--magnitude diagrams characteristic of Milky Way open
clusters pose a fundamental challenge to our traditional understanding
of star clusters as `single stellar populations.' The clear MS
bifurcation of early-type stars with masses greater than
$\sim\unit[1.6]{M_\odot}$ is thought to result from a range in the
stellar rotation rates. However, direct evidence connecting double MSs
to stellar rotation properties has yet to emerge. Here, we show
through analysis of the projected stellar rotational velocities
($v\sin i$, where $i$ represents the star's inclination angle) that
the well-separated double MS in the young, $\sim\unit[200]{Myr}$-old
Milky Way open cluster NGC 2287 is tightly correlated with a
dichotomous distribution of stellar rotation rates. We discuss
  whether our observations may reflect the effects of tidal locking
  affecting a fraction of the cluster's member stars in stellar binary
  systems. We show that the slow rotators could potentially be
  initially rapidly rotating stars that have been slowed down by tidal
  locking by a low mass-ratio companion in a cluster containing a
  large fraction of short-period, low-mass-ratio binaries. This
demonstrates that stellar rotation drives the split MSs in young,
$\lessapprox \unit[300]{Myr}$-old star clusters. However, special
  conditions, e.g., as regards the mass-ratio distribution, might be
  required for this scenario to hold.
\end{abstract}

  
\section{Introduction \label{sec:intro}}

Over the last decade, multiple studies have revealed the common
occurrence of bifurcated main sequences (MSs) in young, $\leqslant
\unit[400]{Myr}$-old star clusters in the Large and Small Magellanic
Clouds \citep[LMC and SMC;][]{2015MNRAS.450.3750M,
  2015MNRAS.453.2637D, 2016MNRAS.458.4368M, 2017MNRAS.465.4363M,
  2017ApJ...844..119L,2017MNRAS.467.3628C, 2018MNRAS.477.2640M}. A
split MS was first observed in NGC 1844 by
\citet{2013A&A...555A.143M}. Combined with extended MS turn-off
regions (eMSTOs), which have been discovered in intermediate-age,
$\leqslant\unit[3]{Gyr}$-old LMC and SMC clusters
\citep[e.g.][]{2008ApJ...681L..17M, 2009A&A...497..755M,
  2011ApJ...737....3G}, these split MSs pose a fundamental challenge
to our traditional understanding of star clusters as prototypes of the
canonical `simple' stellar population.

Unlike the multiple MSs found in globular clusters, which are owing to
variations in helium abundance \citep[for a review,
  see][]{2018ARA&A..56...83B}, the split MSs in young clusters are
believed to have an alternative origin
\citep{2015MNRAS.450.3750M}. Based on a comparison of the clusters'
color--magnitude diagrams (CMDs) with stellar evolutionary models,
\citet{2015MNRAS.453.2637D, 2016MNRAS.458.4368M} suggested that the
observed split MSs are consistent with two coeval populations
characterized by different rotation rates: a non-rotating population
of blue MS (bMS) stars and a rapidly rotating population of red MS
(rMS) stars. The presence of a large fraction of fast rotators in
these clusters was first confirmed by employing Be stars
\citep[rapidly rotating B-type stars exhibiting H$\alpha$
  emission][]{2013A&ARv..21...69R} as probes. Using H$\alpha$ images
observed with the {\sl Hubble Space Telescope} ({\sl HST}),
\citet{2017MNRAS.465.4795B, 2018MNRAS.477.2640M} searched for Be stars
in young, massive LMC clusters. They confirmed that Be stars populate
the rMS and the reddest part of the eMSTO. Spectroscopic evidence
directly confirming the stellar rotation properties was subsequently
provided by \citet{2017ApJ...846L...1D, 2018AJ....156..116M},
supporting this rotation-dominant picture.

Based on the \textit{Gaia} Data Release 2
\citep[DR2;][]{2016A&A...595A...2G, 2018A&A...616A...1G},
\citet{2018ApJ...869..139C} found that eMSTOs are a common feature in
young and intermediate-age clusters in the Milky Way and that they may
be governed by a similar mechanism as in Magellanic Clouds. The
presence of stars with rotation rates ranging from slow to nearly
critical rotation has been reported based on observations of field
stars and open clusters \citep{2006ApJ...648..580H,
  2010ApJ...722..605H}, and the effects of stellar rotation in the
MSTO region have also been studied \citep{2015ApJ...807...58B,
  2015ApJ...807...25B, 2015ApJ...807...24B}. However, a direct
connection between stellar rotation and the presence of a split
MS/eMSTO has only been established for a limited number of
clusters. \citet{2018MNRAS.480.3739B}, \citet{2018ApJ...863L..33M},
and \citet{2019ApJ...876..113S} measured the rotational velocities of
MS stars in three Galactic open clusters (OCs) and confirmed the
presence of populations with different rotation rates.

In this paper, we analyze the Galactic OC NGC 2287, which shows a
clearly split MS following stellar membership selection based on
\textit{Gaia} DR2 \citep{2018ApJ...869..139C}. We verify that the
split MS is a genuine characteristic and not caused by differential
extinction. The bMS and rMS are well separated in projected rotational
velocity $v\sin i$, in the sense that bMS stars are composed of slow
rotators and rMS stars are mainly rapid rotators. MSTO stars also show
a similar correlation between their colors and the projected
rotational velocities. Our results suggest that they follow a
dichotomous distribution in equatorial rotational velocity. To interpret our results, the scenario proposed by \citet{2017NatAs...1E.186D}, which suggests that slowly rotating stars could have been initially rapidly rotating stars that have been slowed down by tidal locking by a low mass-ratio companion, is not ruled out. However, these binaries may, in fact, be subject to
  special conditions in terms of their period and mass-ratio
  distributions.

This article is organized as follows. We present the membership
determination, spectroscopic data, and stellar classification in
Section \ref{sec:data}. In Section \ref{sec:rotation} we analyze the
stellar rotation rates in the split MS and MSTO regions. Our
discussion and conclusions are covered in Section
\ref{sec:discussion}.

\section{Data and Analysis \label{sec:data}}
\subsection{Membership determination}

\begin{figure*}[ht]
\gridline{\fig{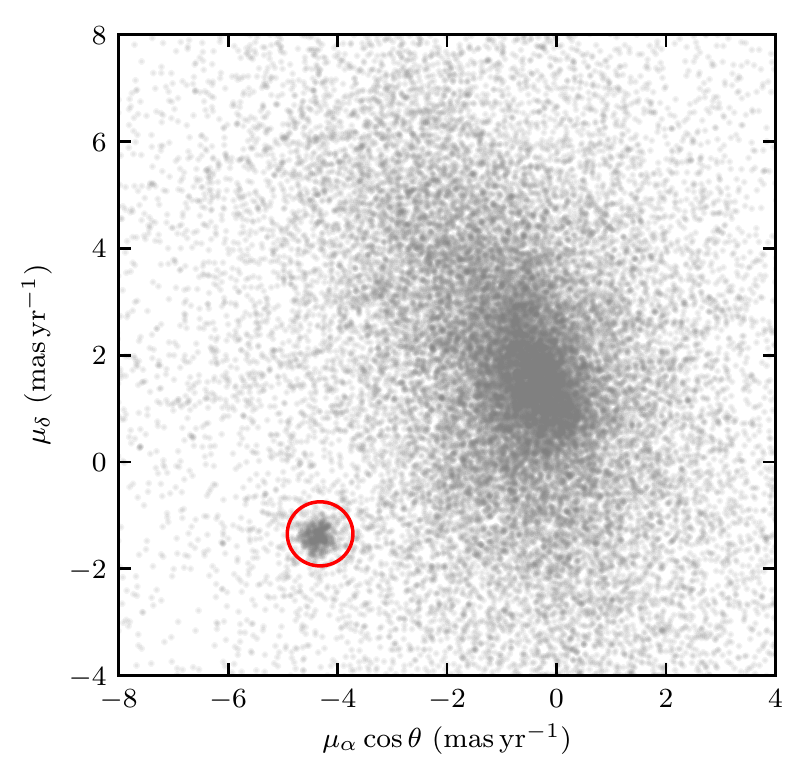}{0.33\textwidth}{}
          \fig{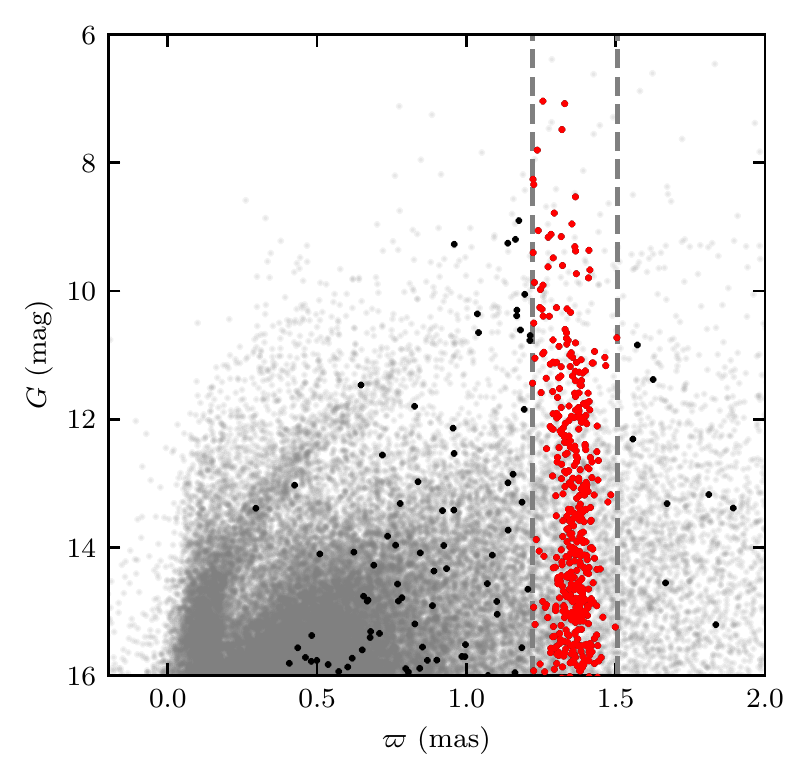}{0.33\textwidth}{}
          \fig{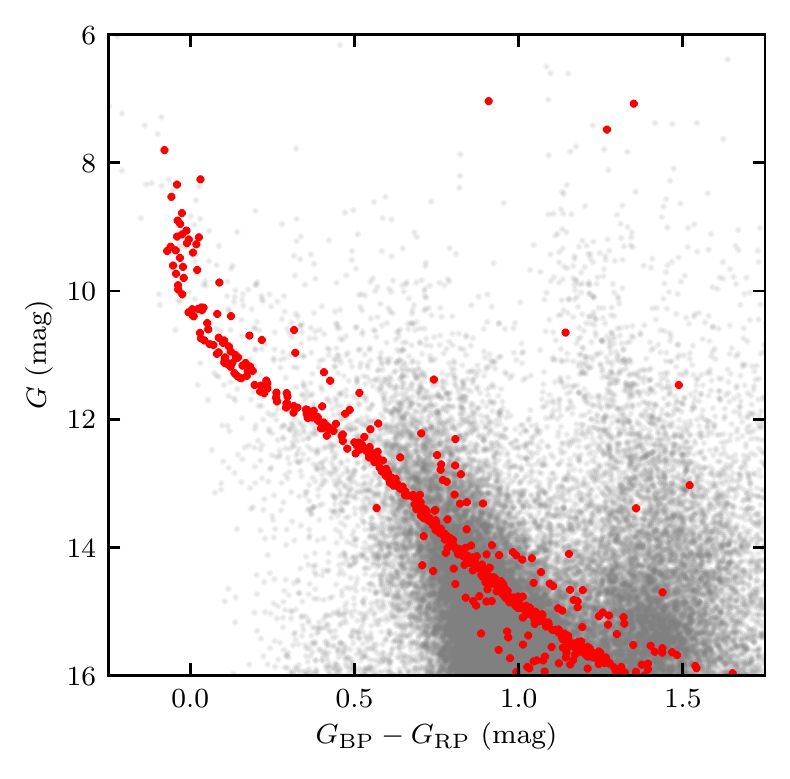}{0.33\textwidth}{}
}
\caption{(left) Vector-point diagram of the proper motions of stars
  brighter than $G = \unit[16]{mag}$ located within 2\degree of the
  NGC 2287 center. The red circle shows the primary selection area
  ($\unit[0.6]{mas\,yr^{-1}}$). (middle) $G$-band photometry versus
  stellar parallaxes. The parallax-selected members are marked as red
  dots. The vertical dashed lines represent the parallax selection
  boundaries. (right) CMD of all stars in the field (grey dots) and
  the NGC 2287 member stars (red solid dots). A split MS is visible
  around $G\sim\unit[11]{mag}$.\label{fig:member}}
\end{figure*}

From \textit{Gaia} DR2, we obtained the photometric measurements and
derived the membership probabilities of the NGC 2287 cluster stars
based on their proper motions and parallaxes. We searched for all
stars within 2\degree of the center of NGC 2287 in the \textit{Gaia}
DR2 database. In the vector-point diagram of its stellar proper
motions (see the left panel of Fig.~\ref{fig:member}), NGC 2287 is
well separated and concentrated at $(\mu_\alpha\cos\theta, \mu_\delta)
\approx \unit[(-4.34, -1.37)]{mas\,yr^{-1}}$. We defined
$\mu_\mathrm{R} =
\sqrt{(\mu_\alpha\cos\theta-<\mu_\alpha\cos\theta>)^2+(\mu_\delta-<\mu_\delta>)^2}$
and used $\mu_\mathrm{R} \le \unit[0.6]{mas\,yr^{-1}}$ as our primary
selection criterion. Next, we calculated the median value of $\varpi$
and the corresponding r.m.s. ($\sigma$) for stars clustered near
$\varpi\sim\unit[1.4]{mas\,yr^{-1}}$. These stars were further
selected based on their parallax measurements, within
$\unit[2]{\sigma}$, i.e., $\unit[1.2215]{mas\,yr^{-1}} \le \varpi \le
\unit[1.5060]{mas\,yr^{-1}}$ (see the red dashed lines in the middle
panel of Fig.~\ref{fig:member}, corresponding to distances between
\unit[818]{pc} and \unit[664]{pc}). The \textit{Gaia}-based CMD of the
cluster's member stars, combined with that of all stars in the field,
is presented in the right panel of Fig.~\ref{fig:member}. NGC 2287
shows multiple clearly defined MSs at $G$ band magnitudes from
\unit[10.5]{mag} to \unit[11.5]{mag}. A binary sequence above the MS
is clearly visible.

\subsection{Spectroscopic data}

For NGC 2287, high-resolution spectra were obtained with the European
Southern Observatory's Very Large Telescope equipped with the
FLAMES/GIRAFFE spectrograph \citep{2002Msngr.110....1P}, collected as
part of programs 380.D-0161(A) and 380.D-0161(B) (PI Gieles). The
HR14B spectral setup employed offers a nominal resolution of $R\approx
\lambda /\Delta\lambda\approx 29,000$, covering wavelengths from
\unit[5139]{\AA} to \unit[5356]{\AA}. We retrieved
wavelength-calibrated spectra from the ESO archive, which had been
reduced by the GIRAFFE pipeline. In total, 166 member stars were
observed as part of these programs. In this paper, we study 53 bright
stars covering the eMSTO, rMS, bMS, and the equal-mass binary sequence
(see Section~\ref{sec:classification} and
Fig.~\ref{fig:cmd_group}). The typical signal-to-noise ratio of the
spectra is $\sim$100. Most of the bright member stars were observed
twice within 3 months.

\subsection{Isochrone fitting}
\begin{figure*}[ht]
\plotone{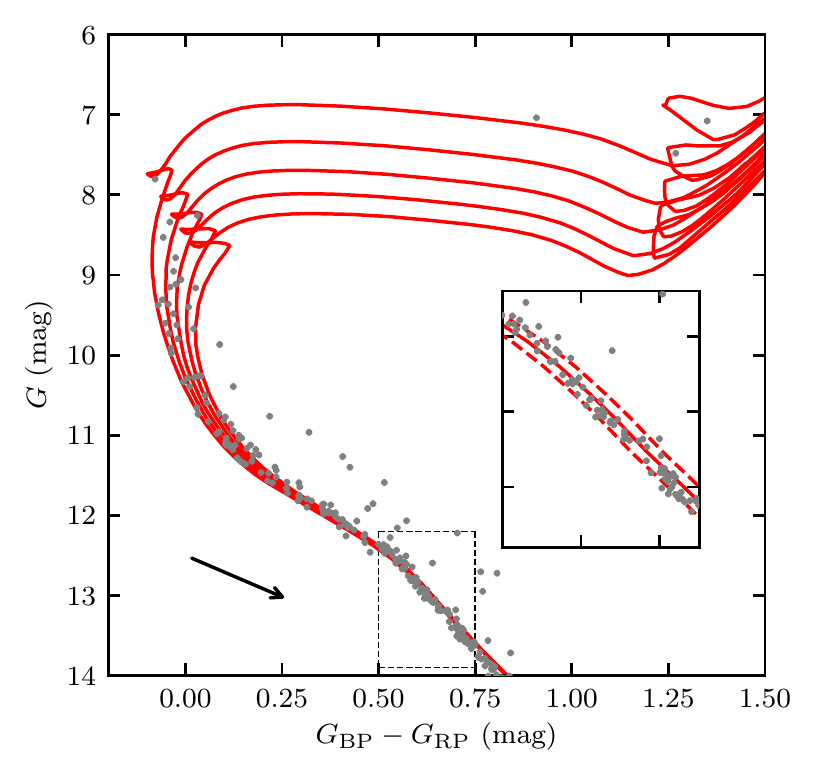}
\caption{CMD of the OC NGC 2287 in the \textit{Gaia} passbands. Grey
  dots represent all stars in the cluster field. Black dots represent
  cluster member stars selected based on \textit{Gaia} DR2. The
  best-fitting (leftmost) PARSEC isochrone \citep{2012MNRAS.427..127B} to the bluest edge of the bulk stellar population has an age
  of \unit[150]{Myr}, a metallicity $Z = 0.0152$, and a distance of
  $\sim\unit[734]{pc}$. Isochrones for ages from \unit[150]{Myr} to
  \unit[350]{Myr} (in steps of \unit[50]{Myr}) are also
  overplotted. The black arrow indicates the direction of the
    reddening vector. The dashed box is shown in more detail in the
  inset, where the solid line is the best-fitting isochrone and the
  dashed lines show the same isochrone shifted by \unit[0.02]{mag} in
  color. \label{fig:cmd}}
\end{figure*}

In Fig.~\ref{fig:cmd}, a set of isochrones from PARSEC 1.2S
\citep{2012MNRAS.427..127B} for solar metallicity $Z = 0.0152$ and
extinction $A_\mathrm{V} = \unit[0.217]{mag}$ is overplotted. The
extinction coefficients for the \textit{Gaia} bands were estimated
based on the \citet{1989ApJ...345..245C} and
\citet{1994ApJ...422..158O} extinction curve with $R_V = 3.1$. The
corresponding ages range from \unit[150]{Myr} to \unit[350]{Myr} in
steps of \unit[50]{Myr}, covering the blue and red ridge lines of the
split MS. We performed our CMD fitting based on visual matching. The
best-fitting isochrone to the blue edge of the bulk stellar
  population has an age of \unit[150]{Myr} and a distance modulus of
9.33 ($\unit[\sim 734]{pc}$).

\citet{2012ApJ...751L...8P} suggested that differential extinction
could generate an artificial eMSTO. However, it is unlikely that the
split MS phenomenon we observe in NGC 2287 is somehow produced by
differential extinction. To prove that the effect of differential
extinction is negligible, we inspected the width of the MS (see the
dashed box in Fig.~\ref{fig:cmd}). In the inset, we show the
best-fitting isochrone (red solid line) as well as that same isochrone
shifted by \unit[0.02]{mag} in color (red dashed lines). We found that
most MS stars are encompassed within the dashed lines, indicating very
limited differential extinction. The clearly isolated equal-mass
binary sequence also suggests that differential reddening is
negligible.

\subsection{Classification \label{sec:classification}}

\begin{figure*}[ht]
\plotone{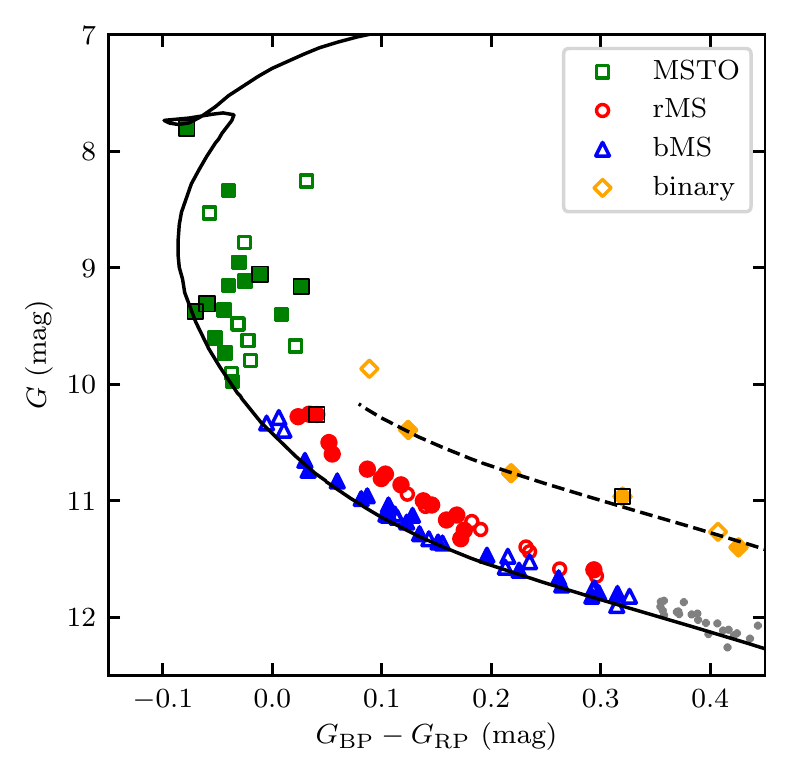}
\caption{CMD of the OC NGC 2287 around its eMSTO and split-MS
  regions. bMS, rMS, MSTO, and binary stars are represented by blue
  triangles, red circles, green squares, and yellow diamonds,
  respectively. Spectroscopically analyzed stars are marked with solid
  markers. The black solid and dashed lines represent the best-fitting
  PARSEC isochrone \citep{2012MNRAS.427..127B} and the corresponding
  equal-mass binary sequence, respectively. Squares represent
  spectroscopic binaries. \label{fig:cmd_group}}
\end{figure*}

Fig.~\ref{fig:cmd_group} shows the CMD pertaining to the cluster's
member stars, in \textit{Gaia} passbands. We classified the member
stars into four groups on the basis on their loci in the CMD: eMSTO
(green squares), bMS (blue triangles), rMS (red circles), and
equal-mass binary systems (yellow diamonds). Stars brighter than the
bulk MS population by \unit[0.75]{mag} were identified as equal-mass
binary stars. Other stars brighter than $G \sim\unit[10]{mag}$ were
classified as eMSTO stars, whereas the fainter stars were assigned to
the bMS and rMS according to their colors. At the faint end, around $G
\sim \unit[12]{mag}$, the split MSs converge. We did not include these
stars in our analysis. The number of stars with available
spectroscopic data account for half the population in each group, and
therefore we estimated that the completeness levels for all groups is
around 50--60\%.

The MS region is composed of three components, including two
well-separated MSs and a sparse sequence populated by equal-mass
binary systems. The (leftmost) bMS can be adequately approximated by a
single-star isochrone, while the parallel and slightly offset rMS is
also well-defined. The discrete nature of the rMS and bMS in the CMD
renders this cluster of particular interest in the context of previous
studies of young clusters, where the target objects were too distant
to perform reliable stellar membership determinations
\citep{2018AJ....156..116M} or where the MS region was smeared out
\citep{2018ApJ...863L..33M}.

We found seven spectroscopic binary candidates in our sample. Only one
is classified as a single-lined spectroscopic binary. Its absorption
lines show a significant shift in radial velocity
($\sim\unit[50]{km\,s^{-1}}$) between two observational visits. The
others are classified as double-lined spectroscopic binaries. The
relevant finding chart did not reveal any suitable companion stars
within \unit[2.0]{arcsec} (\unit[1.2]{arcsec} fiber diameter plus
\unit[0.8]{arcsec} seeing) for these candidates. Most of these
binaries exhibit relatively narrow absorption profiles, except for
\textit{Gaia} ID 2927014822050825856. The latter has a relatively
broad absorption profile accompanied by a second sharp and narrow
absorption line in its center. This additional absorption feature is
only present at wavelengths around \unit[5167]{\AA}. We confirmed that
this feature remained unchanged between two observations separated by
78 days. Since there is no other source discernible close to this
candidate star, the additional absorption in the core of its broad
absorption profile might come from an unresolved binary system
including a secondary star with a low effective temperature but a
rather high metallicity. Such a companion can form a narrow Mg{\sc i}
triplet line without affecting the other components (e.g., the
continuum) of the spectrum. We did not attempt to estimate its $v\sin
i$, because the spectrum cannot be well fitted by a rotationally
broadened profile (see Table~\ref{tab:param}).

\section{Stellar rotation \label{sec:rotation}}

We used the absorption-line profiles of the Mg{\sc i} triplet to
calculate the projected rotational velocities. We first derived
synthetic stellar spectra from the Pollux database
\citep{2010A&A...516A..13P} with effective temperatures
($T_\mathrm{eff}$) from \unit[6400]{K} to \unit[11000]{K} (using steps
of \unit[100]{K}), surface gravities from $\log g = 3.5$ to $\log g =
5.0$ (steps of 0.5) and metallicity from [Fe/H] $=\unit[-1.0]{dex}$ to
[Fe/H] $=\unit[1.0]{dex}$ (steps of \unit[0.5]{dex}). The spectra were
then computed based on the plane-parallel ATLAS12 model atmospheres in
local thermodynamic equilibrium \citep{2005MSAIS...8..189K}. The
microturbulent velocity was fixed at $\unit[2]{km\,s^{-1}}$ for all
models. Synthetic spectra were then generated using the SYNSPEC tool
\citep{1992A&A...262..501H}. A Gaussian kernel of $\sigma =
\unit[0.075]{\AA}$ (corresponding to a spectral resolution of 29,000
at \unit[5100]{\AA}) was applied to the model spectra to represent the
effects of instrumental broadening. Next, they were convolved with a
rotational $v\sin i$ profile from $\unit[5]{km\,s^{-1}}$ to
$\unit[400]{km\,s^{-1}}$ (steps of $\unit[5]{km\,s^{-1}}$) and shifted
by a radial velocity from $\unit[0]{km\,s^{-1}}$ to
$\unit[50]{km\,s^{-1}}$ in steps of $\unit[2]{km\,s^{-1}}$ \citep[the
  average radial velocity is
  $23.3\,\mathrm{km\,s^{-1}}$;][]{2005A&A...438.1163K}. Finally, the
model spectra were resampled to match the observed spectra using
$\chi^2$ minimization.

Since no Balmer lines are covered by the observed wavelength range, it
is hard to determine a reliable effective temperature. Therefore, we
adopted the measured $T_\mathrm{eff}$ from \textit{Gaia} DR2 as our
initial guess for each individual spectrum and searched for the best
model with an effective temperature within \unit[500]{K} of the
initial value. Although the \textit{Gaia} DR2 $T_\mathrm{eff}$
  values have not been corrected for extinction
  \citep{2018A&A...616A...8A}, this is nevertheless a generally
  acceptable practice. On the one hand, the extinction to NGC 2287
  ($A_\mathrm{V} = \unit[0.217]{mag}$) is smaller than the median
  extinction value (\unit[0.335]{mag}) of the low-extinction data set
  used to estimate the effective temperatures
  \citep{2018A&A...616A...8A}, thus allowing us to infer reliable
  effective temperatures of this cluster's member stars using
  $T_\mathrm{eff}$ values from \textit{Gaia} DR2. On the other hand,
  we have also adopted a \unit[500]{K} buffer area to avoid any
  systematic differences from the actual values. The projected
rotational velocity is weakly dependent on some stellar parameters. We
estimated the uncertainty in $v\sin i$ by remeasuring mock spectra
generated from the model template with different signal-to-noise
ratios. This procedure was repeated 100 times and the 68th percentile
of the projected rotational velocity distribution was taken as the
uncertainty. The minimum value of the uncertainty was fixed at
$\unit[5]{km\,s^{-1}}$, reflecting the limitation owing to the grid's
step sizes. As for stars with multiple observations, we confirmed that
there were no significant variations in their stellar parameters,
including in their radial and projected rotational velocities
(exceptions were discussed in the previous section) between each
visit. The typical difference is within the prevailing
uncertainties. The inferred $v\sin i$ values are listed in
Table~\ref{tab:param}.

\startlongtable
\begin{deluxetable*}{lCCCCCc}
\tablecaption{Projected rotational velocities of member stars in NGC 2287.\label{tab:param}}
\tablewidth{0pt}
\tablehead{
\colhead{\textit{Gaia} ID}&\colhead{$G$ (mag)}&\colhead{$G_\mathrm{bp}$ (mag)}&\colhead{$G_\mathrm{rp}$ (mag)} & \colhead{$\Delta G_\mathrm{BP} - G_\mathrm{RP}$} & \colhead{$v\sin i$ ($\unit{km\,s^{-1}}$)\tablenotemark{a}} & \colhead{Classification}  \\
\colhead{(1)} & \colhead{(2)} &
\colhead{(3)} & \colhead{(4)} & \colhead{(5)} & \colhead{(6)} & \colhead{(7)}
}
\startdata
2927207133504349056 & 11.47 & 11.53 & 11.34 & 0.02 & $70 \pm 9$ & bMS \\
2926993995745289216 & 11.66 & 11.75 & 11.49 & 0.03 & $80 \pm 14$ & bMS \\
2926994103128768768 & 11.79 & 11.91 & 11.59 & 0.03 & $50 \pm 15$ & bMS \\
2926990048679648000 & 11.79 & 11.89 & 11.59 & 0.02 & $85 \pm 15$ & bMS \\
2927014787691101696 & 11.28 & 11.33 & 11.19 & 0.01 & $120 \pm 12$ & bMS \\
2927008224981157760 & 10.83 & 10.85 & 10.79 & 0.01 & $60 \pm 13$ & bMS \\
2927199505642285824 & 11.59 & 11.67 & 11.45 & 0.01 & $85 \pm 11$ & bMS \\
2927007262908564736 & 11.12 & 11.16 & 11.06 & 0.01 & $125 \pm 10$ & bMS \\
2927204517860633344 & 10.98 & 11.02 & 10.94 & 0.01 & $130 \pm 8$ & bMS \\
2926995030833789056 & 11.35 & 11.40 & 11.25 & 0.01 & $105 \pm 13$ & bMS \\
2927002040228145280 & 10.96 & 10.99 & 10.90 & 0.02 & $172 \pm 10$ & bMS \\
2926916965512116864 & 11.12 & 11.17 & 11.05 & 0.03 & $215 \pm 11$ & bMS \\
2927013482021158656 & 10.74 & 10.75 & 10.72 & 0.00 & $70 \pm 9$ & bMS \\
2927208336094874624 & 11.04 & 11.07 & 10.97 & 0.02 & $177 \pm 8$ & bMS \\
2927020663206313472 & 11.82 & 11.92 & 11.63 & 0.00 & $160 \pm 18$ & bMS \\
2926989979960168448 & 11.18 & 11.23 & 11.11 & 0.01 & $122 \pm 12$ & bMS \\
2927018734755286784 & 11.72 & 11.82 & 11.55 & 0.01 & $97 \pm 14$ & bMS \\
2927016574388824064 & 10.65 & 10.66 & 10.63 & 0.00 & $185 \pm 9$ & bMS \\
2927213004717564800 & 10.50 & 10.52 & 10.47 & 0.04 & $252 \pm 7$ & rMS \\
2927008980887462400 & 10.26 & 10.28 & 10.24 & 0.06 & $40 \pm 7$ & rMS\tablenotemark{b} \\
2927199849239653760 & 10.73 & 10.76 & 10.68 & 0.05 & $290 \pm 8$ & rMS \\
2927203113414681600 & 11.00 & 11.05 & 10.91 & 0.06 & $290 \pm 8$ & rMS \\
2927022003236076800 & 11.33 & 11.39 & 11.21 & 0.03 & $205 \pm 13$ & rMS \\
2926911223137755136 & 10.28 & 10.29 & 10.27 & 0.04 & $277 \pm 29$ & rMS \\
2926917137310902016 & 10.60 & 10.62 & 10.56 & 0.04 & $252 \pm 7$ & rMS \\
2926994824683234944 & 10.77 & 10.81 & 10.71 & 0.06 & $80 \pm 9$ & rMS \\
2927199746160410496 & 11.16 & 11.22 & 11.06 & 0.05 & $247 \pm 10$ & rMS \\
2927012657387390336 & 10.26 & 10.27 & 10.24 & 0.05 & $277 \pm 8$ & rMS \\
2927008431139550080 & 11.59 & 11.70 & 11.40 & 0.08 & $100 \pm 12$ & rMS \\
2926914182373505152 & 11.12 & 11.19 & 11.02 & 0.07 & $220 \pm 9$ & rMS \\
2926912533106020096 & 10.86 & 10.91 & 10.79 & 0.06 & $275 \pm 8$ & rMS \\
2926916175231589888 & 11.04 & 11.09 & 10.94 & 0.06 & $90 \pm 8$ & rMS \\
2926912361307334528 & 10.81 & 10.85 & 10.75 & 0.05 & $232 \pm 8$ & rMS \\
2927016368239015296 & 11.25 & 11.32 & 11.14 & 0.05 & $250 \pm 11$ & rMS \\
2927013791258742912 & 9.60 & 9.59 & 9.64 & 0.01 & $108 \pm 6$ & MSTO \\
2927209057649389696 & 9.31 & 9.29 & 9.35 & 0.02 & $16 \pm 12$ & MSTO\tablenotemark{b} \\
2927213112098494592 & 9.38 & 9.36 & 9.43 & 0.00 & $18 \pm 11$ & MSTO\tablenotemark{b} \\
2927010905040796288 & 8.95 & 8.95 & 8.98 & 0.05 & $247 \pm 5$ & MSTO \\
2927005682360625664 & 8.34 & 8.33 & 8.37 & 0.04 & $197 \pm 5$ & MSTO \\
2927203388292592384 & 9.73 & 9.72 & 9.76 & 0.01 & $280 \pm 5$ & MSTO \\
2926993175415844096 & 9.98 & 9.97 & 10.01 & 0.00 & $92 \pm 7$ & MSTO \\
2927009255773305856 & 9.40 & 9.41 & 9.40 & 0.08 & $322 \pm 6$ & MSTO \\
2927012515646285440 & 7.80 & 7.78 & 7.86 & 0.00\tablenotemark{c} & $24 \pm 10$ & MSTO\tablenotemark{b} \\
2927006850591725568 & 9.06 & 9.06 & 9.07 & 0.07 & $20 \pm 11$ & MSTO\tablenotemark{b} \\
2927002723118812928 & 9.36 & 9.36 & 9.40 & 0.03 & $157 \pm 6$ & MSTO \\
2927220465082571776 & 9.15 & 9.14 & 9.18 & 0.04 & $235 \pm 5$ & MSTO \\
2927014822050825856 & 9.16 & 9.17 & 9.15 & 0.11 & ... & MSTO\tablenotemark{b} \\
2927014100496388736 & 9.12 & 9.11 & 9.14 & 0.06 & $270 \pm 5$ & MSTO \\
2927008396779815808 & 11.59 & 11.78 & 11.27 & 0.30 & $155 \pm 13$ & binary \\
2927004754647430784 & 10.76 & 10.85 & 10.63 & 0.18 & $107 \pm 10$ & binary \\
2927015200007933568 & 10.39 & 10.44 & 10.32 & 0.13 & $75 \pm 8$ & binary \\
2927019529334958848 & 10.96 & 11.09 & 10.77 & 0.25 & $18 \pm 11$ & binary\tablenotemark{b} \\
2927007950103249920 & 11.40 & 11.56 & 11.13 & 0.27 & $37 \pm 12$ & binary \\
\enddata
\tablenotetext{a}{Uncertainty estimated from the mock test.}
\tablenotetext{b}{Likely a spectroscopic binary.}
\tablenotetext{c}{This star is located at the tip of the blue edge,
  where the isochrone does not fit well. Thus, the pseudo-color for
  this star is fixed at 0 mag.}
\tablecomments{(1) \textit{Gaia} DR2 ID; (2, 3, 4) \textit{Gaia}
  bands; (5) Pseudo-color; (6) Projected rotational velocity; (7)
  Classification based on CMD.}
\end{deluxetable*}

\begin{figure*}[ht]
\plottwo{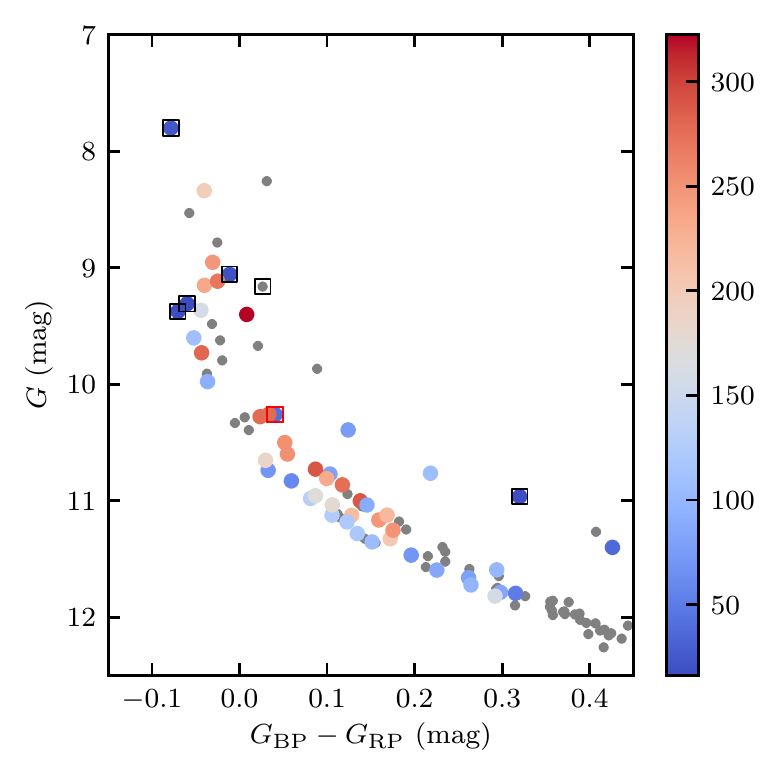}{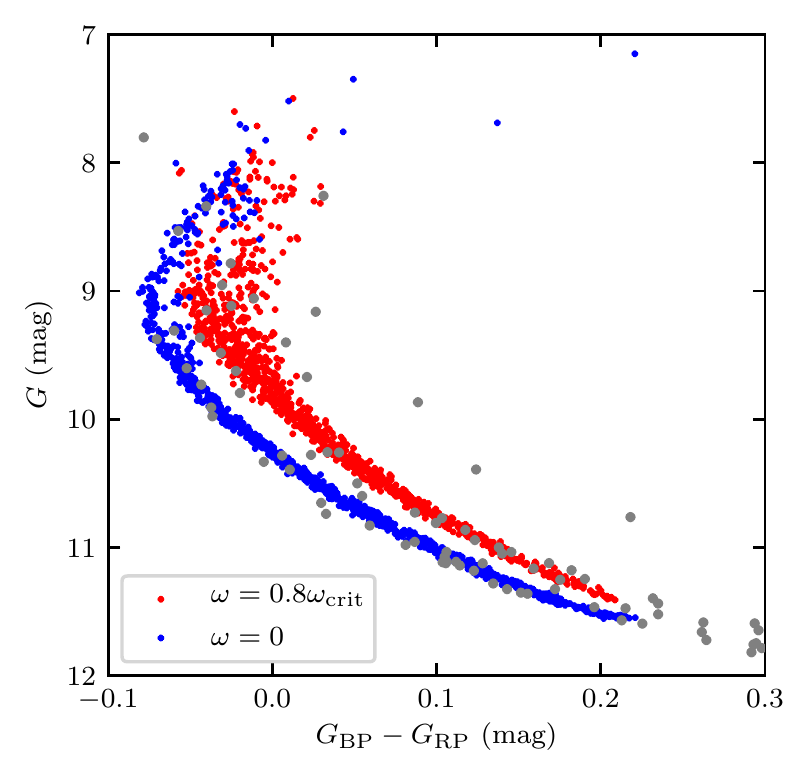}
\caption{(left) CMD of NGC 2287 with stars color-coded according to
  their projected rotational velocities. Squares represent
  spectroscopic binaries. \textit{Gaia} ID 2927008980887462400 is
  indicated by the red square. (right) Synthetic clusters of slowly
  (blue) and rapidly rotating populations (red;
  $\omega=0.8\omega_\mathrm{crit}$) at an age of \unit[150]{Myr}. The
  expected photometric errors have been added to the
  simulations.\label{fig:cmd_rot}}
\end{figure*}

The $v\sin i$ values for 53 bright cluster members around the eMSTO,
bMS, rMS, and the equal-mass binary sequence span a wide range, from
non-rotators to rotation rates greater than
$\unit[300]{km\,s^{-1}}$. The left panel of Fig.~\ref{fig:cmd_rot}
shows the CMD of NGC 2287 with its member stars color-coded by their
projected rotational velocities. bMS and rMS stars are well-separated
in projected rotational velocity: bMS stars are mainly composed of
slow rotators while rMS stars are predominantly rapid rotators. MSTO
stars follow a similar trend, with stars getting redder as their
rotation rates increase; the cluster's equal-mass binary stars are all
slow rotators.

To provide a direct comparison, we overplotted two synthetic, coeval
clusters ($t=\unit[150]{Myr}$) with different rotational rates
($\omega = 0$ and $\omega = 0.8\omega_\mathrm{crit}$) in the right
panel of Fig.~\ref{fig:cmd_rot}. The simulation was carried on based
on SYCLIST models \citep{2013A&A...553A..24G, 2014A&A...566A..21G},
assuming a metallicity of $Z = 0.014$. The model took into account
limb darkening \citep{2000A&A...359..289C}, and adopted the
gravity-darkening law of \citet{2011A&A...533A..43E}. We also adopted
a flat distribution of rotation rates, using single stars only (no
binary systems). Realistic photometric uncertainties of
\unit[0.003]{mag} in both $G$ and $G_\mathrm{BP} - G_\mathrm{RP}$ were
added to the simulation. Both the split MS and eMSTO can be well
reproduced by these synthetic clusters, favoring a rotation-dominant
scenario.

\begin{figure*}[ht]
\plotone{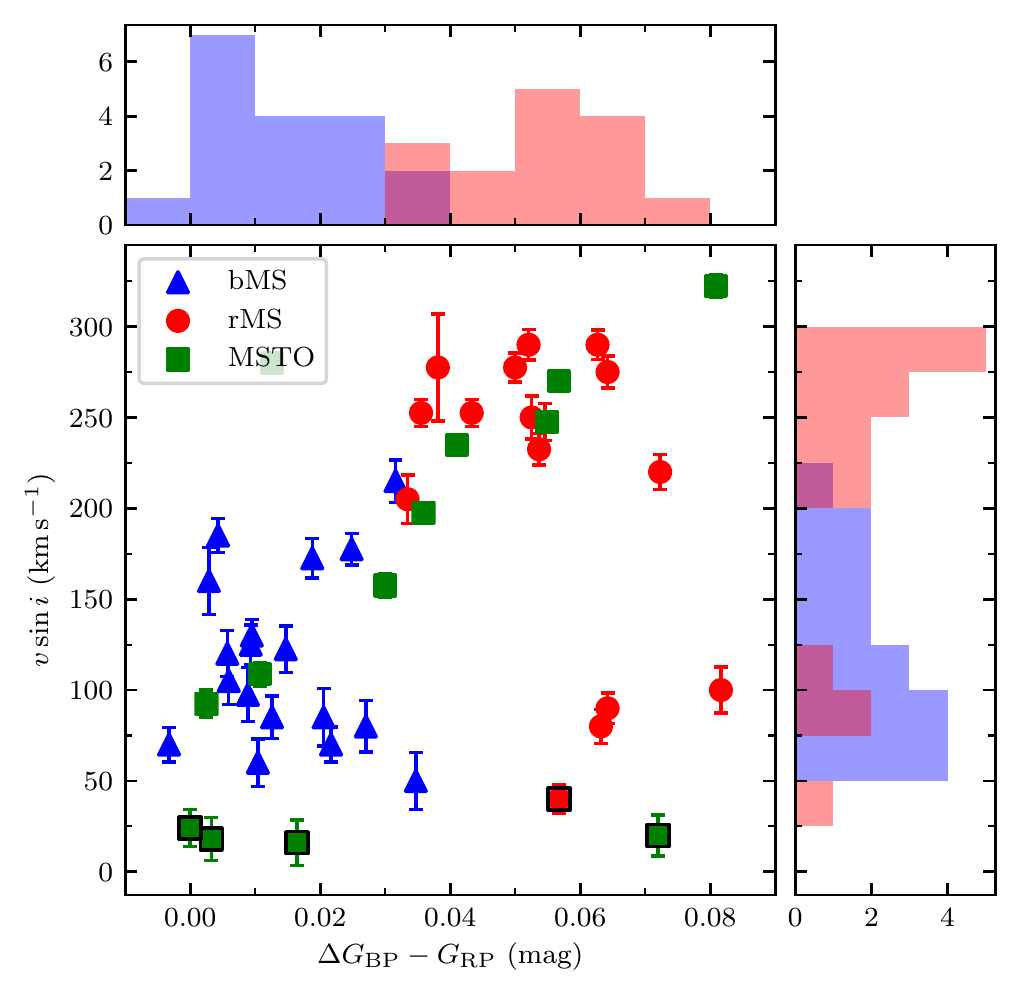}
\caption{Correlation between stellar rotation rates ($v\sin i$) and
  their CMD loci (represented by pseudo-color, $\Delta G_\mathrm{BP} -
  G_\mathrm{RP}$) for split-MS and MSTO stars. The rMS, bMS, and MSTO
  stars are represented by red solid circles, blue triangles, and
  green squares, respectively. The top and right panels show
  histograms of the pseudo-color and $v\sin i$ distributions of the
  bMS and rMS stars, respectively. \label{fig:pseudo}}
\end{figure*}

We now define the pseudo-color $\Delta G_\mathrm{BP} - G_\mathrm{RP}$,
that is, a star's color difference with respect to the cluster's blue
ridge line, represented by the best-fitting isochrone to the cluster's
dominant stellar population (see
Table~\ref{tab:param}). Fig.~\ref{fig:pseudo} shows the distribution
of the NGC 2287 split-MS stars in the $\Delta G_\mathrm{BP} -
G_\mathrm{RP}$ versus $v\sin i$ diagram, as well as their respective
histograms. The majority of cluster member stars exhibit a clear
correlation between $v\sin i$ and pseudo-color, in the sense that the
bMS and rMS populations tend to represent slow and rapid rotational
velocities, respectively. The mean projected rotational velocity for
bMS stars is $\langle v\sin i\rangle_\mathrm{bMS} =
\unit[111\pm13]{km\,s^{-1}}$ ($\sigma = \unit[46]{km\,s^{-1}}$).

Four rMS stars have projected rotational velocities below
$\unit[100]{km\,s^{-1}}$; these coincide with bMS stars in projected
velocity space. One of these objects, \textit{Gaia} ID
2927008980887462400 (indicated by the red square in
Fig.~\ref{fig:cmd_rot}), is a double-lined spectroscopic binary
system. It exhibits an additional absorption profile at a wavelength
around \unit[5178]{\AA}, which may be related to the presence of a
fainter and cooler companion star. Under this assumption and since the
primary star's profile is not significantly affected by the companion,
we adopted the projected rotational velocity given by the best-fitting
model spectrum as a reliable estimate of the primary star's rotation
rate. This system's locus on the rMS is therefore determined by its
binary nature. The binary's primary star is expected to be located on
the bMS, but it appears redder because of contamination by its
unresolved lower-mass (and thus redder) companion. Unresolved binaries
will appear brighter and redder than the (single) primary-star MS,
thus potentially projecting bMS stars onto the rMS
\citep{2013MNRAS.436.1497L}. Three other slowly rotating rMS objects
may also be binary systems. Time-domain photometry would be required
to ascertain their binary nature.

Except for these four objects, all 11 other rMS stars have rotational
velocities well in excess of those characteristic of the bMS
stars. The mean projected rotational velocity of rMS stars is then
$\langle v\sin i\rangle_\mathrm{bMS} = \unit[255\pm10]{km\,s^{-1}}$
($\sigma = \unit[26]{km\,s^{-1}}$). The Spearman coefficient for the
correlation between the pseudo-colors and $v\sin i$ is 0.68 for both
the bMS and the rMS ($p=2.8\times10^{-5}$). The cluster's eMSTO stars
show a similar pattern, in the sense that rapidly rotating stars are
generally redder than slowly or non-rotating stars. These results
strongly suggest that stellar rotation is the underlying physical
cause of the split MS and the eMSTO in NGC 2287.

NGC 2287 also exhibits a slightly extended MSTO region (see
Fig.~\ref{fig:cmd}), which is a common feature of all massive clusters
(with masses greater than about $\unit[10^4]{M_\odot}$) in the
Magellanic Clouds younger than about 3 Gyr and likely caused by
stellar rotation \citep{2014Natur.516..367L}. Through analysis of the
eMSTO's morphology, we conclude that the likely effects of stellar
rotation in the NGC 2287 CMD may instead be incorrectly interpreted as
an age spread spanning $\sim$100 Myr. This is consistent with previous
studies of massive Magellanic Cloud clusters
\citep{2015MNRAS.453.2070N} and open clusters in the Milky Way
\citep{2018ApJ...869..139C}. Note that the split between the bMS and
rMS does not extend as far as the eMSTO region. This is mainly owing
to the compound effects of gravity darkening and rotational
mixing. Whereas gravity darkening associated with rapid rotation
causes stars to have apparently lower surface temperatures
\citep{1924MNRAS..84..665V}, rotational mixing prolongs the stellar MS
lifetime \citep{2010A&A...509A..72E}, resulting in bluer colors
compared with their slowly rotating counterparts by pulling more fresh
hydrogen into the stellar core \citep{2013A&A...553A..24G}. \citet{2011MNRAS.412L.103G} computed evolutionary tracks and
  isochrones of models with and without rotation. They found that
  although rapid rotation indeed influences the shape of the
  evolutionary tracks, isochrones for rapidly rotating stars trace an
  indistinguishable MSTO locus with respect to their non-rotating
  counterparts owing to their longer MS lifetimes. This mechanism,
which is most prominent around the eMSTO region in NGC 2287 (for
stellar masses $\unit[3]{M_\odot} \le M \le \unit[4]{M_\odot}$), can
offset the effect of gravity darkening, thus explaining the
disappearance of the bifurcation towards brighter luminosities.

\section{Discussion and conclusions\label{sec:discussion}}

Numerous studies of young and intermediate-age (1--3 Gyr-old) star
clusters have enriched our understanding of stellar rotation in the
cluster environment \citep{2017ApJ...846L...1D, 2018MNRAS.480.3739B},
and it has become widely acknowledged that stellar rotation plays an
important role in changing the morphology of the clusters' CMDs. One
of the remaining questions in this field relates to the actual
distribution of equatorial rotational velocities $v_\mathrm{eq}$ and
the inclination angle $i$. The straightforward assumption that the
distribution of stellar rotation axes is stochastic would translate
into a uniform three-dimensional orientation distribution. However, a heated dispute as to whether the stellar spin axes in star
  clusters may be aligned (`spin alignment') is currently playing out
  in the literature. \citet{2017NatAs...1E..64C} used asteroseismology
  to measure the inclination angles of the spin axes of 48 red giant
  stars in two open clusters, NGC 6791 and NGC 6819, and reported
  strong stellar-spin alignment in both clusters. However,
  \citet{2018A&A...618A.109M} instead demonstrated non-alignment of
  the rotation axes of stars found in these same two clusters. Apart
  from asteroseismology, several studies have used spectroscopic
  rotation measurements ($v\sin i$), as well as information pertaining
  to stellar radii and photometric rotation periods to disentangle the
  influence of rotational velocity and inclination
  angle. \citet{2010MNRAS.402.1380J} and \citet{2018MNRAS.476.3245J}
  found no evidence of strong alignment in the Pleiades and $\alpha$
  Persei star clusters. However, \citet{2018A&A...612L...2K} claimed
  spin alignment in the Praesepe cluster using similar techniques
  \citep[but also see][for a critical
    discussion]{2018MNRAS.479..391K}.

The distribution of projected rotational velocities is an effective
tool to constrain the underlying distribution of $v_\mathrm{eq}$ and
$i$ \citep{2019NatAs...3...76L}. Here, we consider three cases. Case 1
adopts a uniform distribution of both $v_\mathrm{eq}$ and orientation
in three-dimensional space. Case 2 (spin alignment) is characterized
by a uniform distribution of $v_\mathrm{eq}$ and a Gaussian
distribution of $i$. Case 3 considers a uniform distribution of $i$
and a bimodal distribution of $v_\mathrm{eq}$ to represent the slowly
and rapidly rotating populations, $v_\mathrm{s}$ and $v_\mathrm{r}$,
respectively. The relative numbers of both stellar populations were
carefully considered.

In the simulation, the maximum $v_\mathrm{eq}$ was fixed at
$\unit[414]{km\,s^{-1}}$, corresponding to the maximum
$v_\mathrm{crit}$ in our synthetic clusters ($t=\unit[150]{Myr}$). A
stochastic orientation in three-dimensional space can be represented
by a uniform distribution in $\cos i$
\citep{2019NatAs...3...76L}. Case 2 has two free parameters to
describe its Gaussian profile---the peak inclination angle
$i_\mathrm{peak}$ (5\degree to 90\degree in steps of 5\degree) and the
dispersion $\sigma_i$ (1\degree to 46\degree in steps of
5\degree). For Case 3, $v_\mathrm{s}$ and $v_\mathrm{r}$ range from
$\unit[80]{km\,s^{-1}}$ to $\unit[200]{km\,s^{-1}}$ and from
$\unit[200]{km\,s^{-1}}$ to $\unit[400]{km\,s^{-1}}$, respectively,
both in steps of $\unit[20]{km\,s^{-1}}$. We confirmed that changing
the velocity dispersion did not significantly affect the result, and
thus we fixed the velocity dispersion at $\unit[30]{km\,s^{-1}}$ for
convenience. The number fraction of rapid rotators with respect to the
entire population ranges from 0.1 to 0.9 in steps of 0.1. For each
configuration, we generated 1000 samples and compared them with the
observed results using a two-sample Kolmogorov--Smirnov test. To
reduce stochastic effects, we repeated this process 1000 times for
each configuration and used the median $p$ value to select which
configuration had the highest probability to reproduce the observed
distribution.

\begin{figure*}[ht]
\plotone{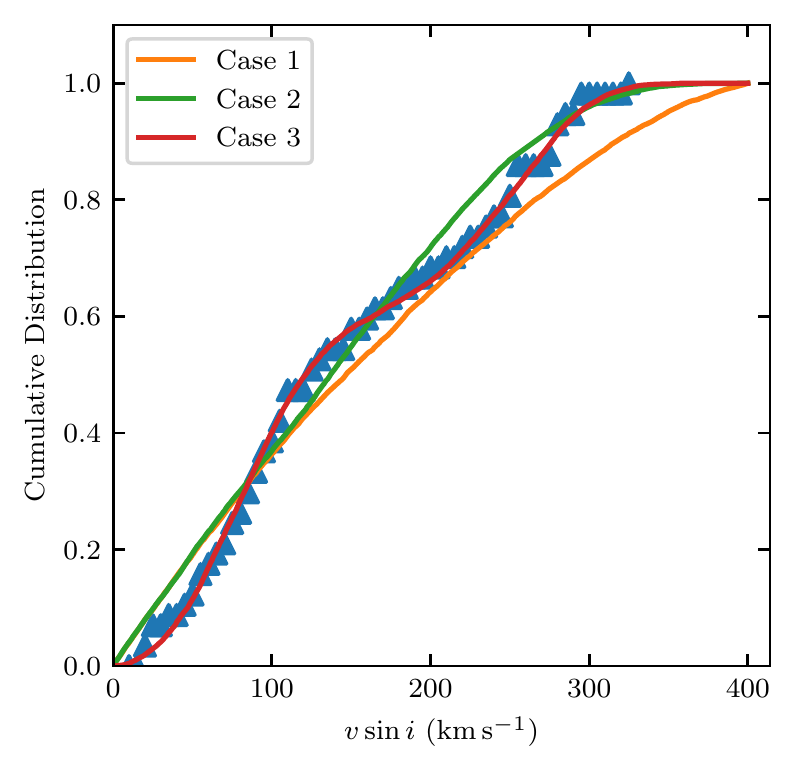}
\caption{Cumulative $v\sin i$ distribution and best-fitting
  theoretical models. The observational data are represented by blue
  triangles. We adopted a maximum $v_\mathrm{eq} =
  \unit[414]{km\,s^{-1}}$. Case 1 (orange): Uniform distributions of
  both $v_\mathrm{eq}$ and $i$. Case 2 (green): Uniform distribution
  of $v_\mathrm{eq}$, combined with a Gaussian distribution of
  $i$. Case 3 (red): Bimodal distribution of $v_\mathrm{eq}$, combined
  with a uniform distribution of $i$.\label{fig:cumulative}}
\end{figure*}

\begin{figure*}[ht]
\plotone{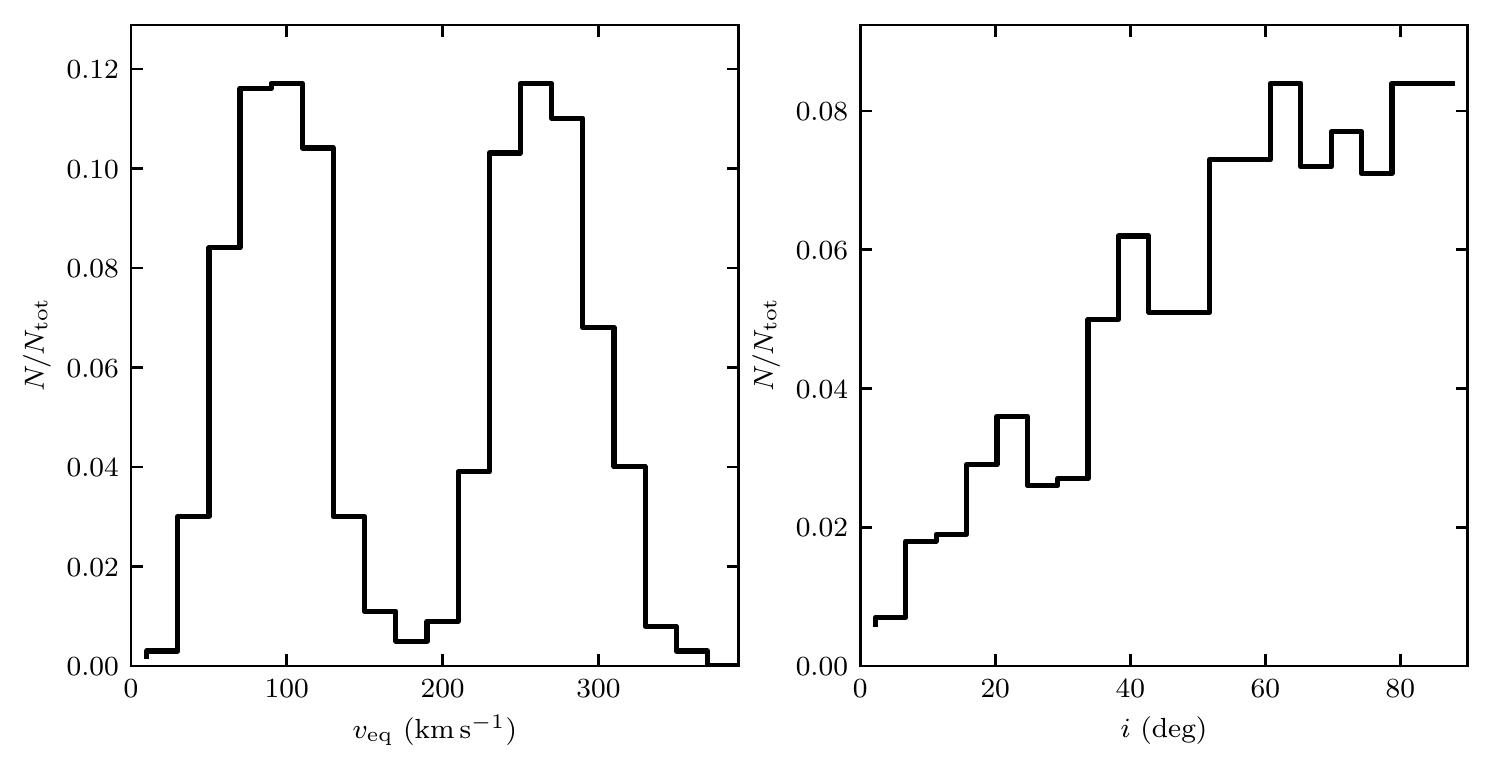}
\caption{Example of best-fitting $v_\mathrm{eq}$ and $i$ distributions
  for Case 3. The bimodal distribution of velocities peaks at both
  $\unit[100]{km\,s^{-1}}$ and $\unit[280]{km\,s^{-1}}$. The velocity
  dispersion is $\unit[30]{km\,s^{-1}}$ and the number ratio of the
  slowly and rapidly rotating populations is
  unity. \label{fig:distribution}}
\end{figure*}

Fig.~\ref{fig:cumulative} shows the cumulative distributions of the
projected rotational velocities, as well as the best-fitting
simulation results representing Cases 1 to 3. Cases 1 and 2 clearly
cannot reproduce the observations (the best-fitting parameters for
Case 2 are $i_\mathrm{peak} = 45\degree$; $\sigma_i=11\degree$). The
best-fitting model of Case 3 ($p = 0.99$; better than Case 2 at the
$\unit[3]{\sigma}$ level) yields two rotating populations with peak
velocities of $v_\mathrm{r} = \unit[280]{km\,s^{-1}}$ and
$v_\mathrm{s} = \unit[100]{km\,s^{-1}}$. This is indeed within
$\unit[1]{\sigma}$ of $\langle v\sin i\rangle_\mathrm{rMS}$ and
$\langle v\sin i\rangle_\mathrm{bMS}$. The number ratio of the two
populations is $n_\mathrm{r}/n_\mathrm{s} = 1$ (see also
Fig.~\ref{fig:distribution}). This represents the first evidence in
support of a dichotomous distribution of real rotational velocities in
star clusters. It raises an important question as to the origin of
such a bimodal rotational velocity distribution.

In fact, this is not the first time we see such a bimodal
  rotation distribution. In a census of some 1100 early-type field
  stars, \citet{2007A&A...463..671R} found that early-type MS stars
  show genuine bimodal distributions of true equatorial rotational
  velocities. Subsequent work by \citet{2012A&A...537A.120Z} confirmed
  that stars more massive than $\unit[2.5]{M_\odot}$ exhibit a bimodal
  equatorial velocity distribution, while less massive stars show a
  unimodal rotation distribution. However, note that the stellar mass
  range of the bimodal rotation distribution reported by
  \citet{2012A&A...537A.120Z} is not exactly the same as the masses
  relevant for the split MS and MSTO observed for NGC 2287. The mass
  of the split MS's faintest end is estimated at around
  $\unit[1.7]{M_\odot}$, which is significantly lower than the
  dividing mass of the bimodal and unimodal rotation distribution
  found by \citet{2012A&A...537A.120Z}. Another difference is found in
  the rotational velocities for the slow and rapid rotators. The peak
  rotational velocities we derived for NGC 2287 are slightly larger
  than the values reported by \citet{2012A&A...537A.120Z} for a
  similar mass range ($\unit[\sim40]{km\,s^{-1}}$ and
  $\unit[\sim210]{km\,s^{-1}}$ for slow and fast rotators,
  respectively). These disagreements remain even if we consider the
  effects associated with the different evolutionary MS phases for
  different masses.

\citet{2012A&A...537A.120Z} argued that in late A-type stars
  ($\unit[1.6]{M_\odot} \leqslant M \leqslant \unit[2.5]{M_\odot}$),
  rotational velocities are accelerated during the first one-third of
  their MS lifetime and then remain high for a long time, while
  massive stars ($\unit[2.5]{M_\odot} \leqslant M \leqslant
  \unit[3.5]{M_\odot}$) undergo efficient deceleration up to the
  terminal MS age. However, essential evidence is still lacking to
  account for the high fraction of slow rotators in the high-mass
  regime and vice versa. These conditions are unable to explain our
  observations of NGC 2287. \citet{2012A&A...537A.120Z} suggested that
  tidal braking by close binaries could be at work. In fact,
  \citet{2017NatAs...1E.186D} argued that the bMS in young clusters
might be the outcome of fast braking of the rapidly rotating
population. Magnetic-wind braking or tidal torques owing to a binary
companion can, in theory, rapidly decelerate a star's rotation rate
(that is, over a short period compared with the cluster's age) and
transfer the star's evolution from the rapidly rotating to the
non-rotating track. Meanwhile, the core hydrogen content would remain
unchanged, thus making the cluster appear younger in the CMD. Here, we explore whether the bMS stars in NGC 2287 could be composed
  of a population of slow rotators ($\sim \unit[100]{km\,s^{-1}}$)
  which may have slowed down from their initial state of rapid
  rotation by low-mass-ratio ($q \le 0.4$) binary companions.

We adopted $\unit[2]{M_\odot}$ and $\unit[2]{R_\odot}$ as,
respectively, the typical mass and radius of the bMS population and
estimated the synchronization timescale for such a star with a
radiative envelope \citep{2002MNRAS.329..897H},
\begin{equation}
	\frac{1}{\tau_\mathrm{sync}}=52^{5/3}\left(\frac{GM}{R^3}\right)^{1/2}\frac{MR^2}{I}q_2^2(1+q_2)^{5/6}E_2\left(\frac{R}{a}\right)^{17/2},
\end{equation}
where $G$ is the gravitational constant, $M$ and $R$ the stellar mass
and radius, respectively, $q_2$ the mass ratio, $I$ the moment of
inertia and $E_2$ is a second-order tidal coefficient which can be
fitted to values given by \citet{1975A&A....41..329Z}, i.e.,
\begin{equation}
    E_2=1.592\times10^{-9}M^{2.84}.
\end{equation}

\begin{figure*}[ht]
\centering
\includegraphics{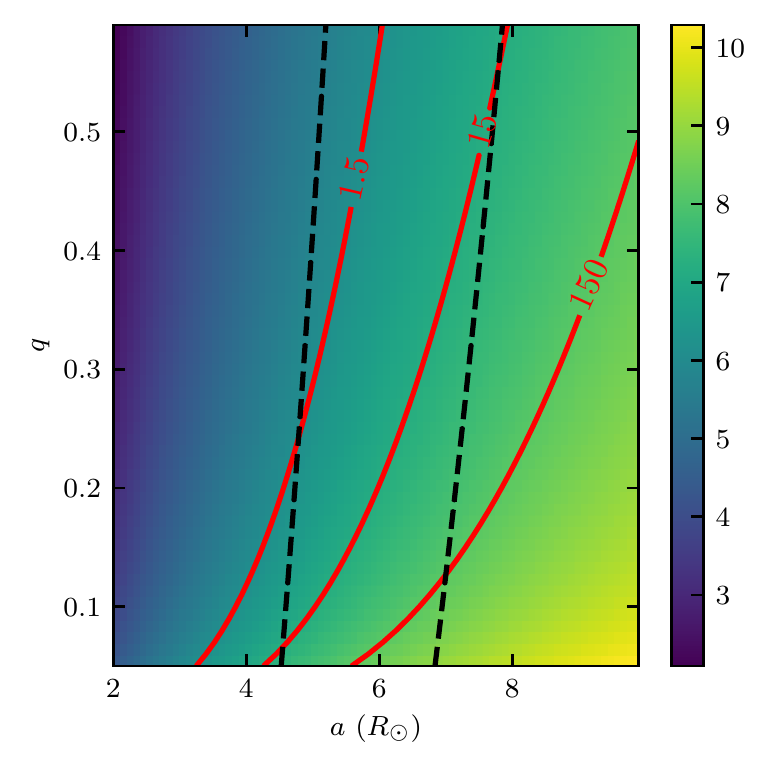}
\caption{Synchronization timescale of a $\unit[2]{M_\odot}$ primary
  star for various mass ratios ($q$) and binary separations ($a$). The
  color represents the synchronization timescale (in logarithmic
  units, yr). The three contour lines represent \unit[1.5]{Myr},
  \unit[15]{Myr} and \unit[150]{Myr} (1\%, 10\%, and 100\% of the
  cluster's age). The black dashed lines enclose the expected relation
  between $a$ and $q$ for slow rotators if we assume that our binary
  systems are already tidally locked and the orbital and rotational
  periods are identical. \label{fig:timescale}}
\end{figure*}

In Fig.~\ref{fig:timescale}, we display the synchronization timescale
for various mass ratios $q$ and binary separations $a$ for a
$\unit[2.0]{M_\odot}$ star. The synchronization timescale of a close
binary system ($a \le \unit[7]{R_\odot}$) is relatively short compared
with the age of NGC 2287. We also overplotted the expected relation
between $a$ and $q$ for slow rotators, calculated using Kepler's Third
Law, assuming that the binary system is already tidally locked and the
orbital and rotational periods are the same. The corresponding orbital
period is around $P=1$ day for $v_\mathrm{eq}=\unit[100]{km\,s^{-1}}$
($\sigma = \unit[30]{km\,s^{-1}}$). The region in the $a$ versus $q$
diagram where such stars may be found overlaps with the
synchronization timescale mainly between \unit[1.5]{Myr} and
\unit[15]{Myr}.

The colors of higher-mass-ratio ($q \ge 0.5$) unresolved binaries will
be reddened significantly compared with those of their
lower-mass-ratio counterparts. Unresolved higher-mass-ratio binary
systems will therefore appear on the rMS rather than the bMS. Tidal
locking, that is, the synchronization of the primary and secondary
stellar rotation rates can effectively drain a system's angular
momentum into the orbital system \citep{2013ApJ...764..166D}. The
synchronization timescale \citep{2002MNRAS.329..897H} for a typical
bMS star in a close binary system is, in general, much shorter than
the host cluster's age, particularly so for NGC 2287 (see
Fig.~\ref{fig:timescale}). By the time the cluster has evolved to its
current age, such close binary systems will have become locked through
tidal interactions, thus forming a population of slow rotators
segregated in color from rapid rotators. Assuming that the binary
system is tidally locked and the orbital period ($a$) is the same as
the rotational period, the expected relation between $a$ and $q$ for
slow rotators also supports the notion that such binary systems are
likely to decelerate the stellar rotation rates and, thus, become a
bMS population as observed.

The scenario advocated by \citet{2017NatAs...1E.186D} cannot be
  ruled out. However, the bMS-to-rMS number ratio of near unity we
  have found in NGC 2287 is hard to explain on this
  basis. \citep{2017ApJS..230...15M} found that, unlike the situation
  for solar-type stars, which are characterized by a log-normal
  orbital-period distribution with a peak near $\log P \approx 5$
  [days], O- and early-B type stars show a bimodal distribution with
  peaks at short ($\log P \le 1$ [days]) and intermediate ($\log P
  \approx 3.5$ [days]) periods. As the primary mass increases, the
  fraction of rapidly orbiting binaries increases and could be
  comparable to or even higher than the fraction of binaries with
  longer orbital periods. However, the fraction of short orbital
  period binaries becomes significant only for masses in excess of
  $\unit[9]{M_\odot}$. Although the reasons are still unclear, this
  may be associated with some mechanism driving the orbital evolution
  which causes the companion to migrate inwards, to the inside of the
  circumstellar disk. However, the bMS-to-rMS number ratio we have
  found for NGC 2287 does not fit easily in this pattern. In addition,
  its mass ratio distribution is also peculiar. A few systems are
located perfectly on the equal-mass binary sequence, implying that
they have mass ratios close to unity. Our observational evidence shows
that in NGC 2287, binary systems have either very low or very high
(near unity) mass ratios, since no objects with intermediate mass
ratios appear to be present in the CMD.

Our results imply that the clearly separated double MSs in NGC 2287
can result from a random distribution of inclination angles and a
dichotomous distribution of equatorial rotational velocities with
peaks near $\unit[100]{km\,s^{-1}}$ and $\unit[280]{km\,s^{-1}}$. The slow rotators are likely stars that initially
rotated rapidly but subsequently slowed down through tidal locking
induced by low-mass-ratio binary systems. The short tidal-braking
timescale characteristic of these bMS stars ($\le$ 10\% of the
cluster's age) thus renders the distribution of rotational velocities
dichotomous. However, the cluster would have a much larger population of short period binaries than is seen in the literature, with relatively low secondary masses.
Although still speculative, future time-domain
observations of bMS stars may shed additional light on the underlying
physical processes.

\acknowledgments R. d. G. and L. D. acknowledge research support from
the National Natural Science Foundation of China through grants
11633005, 11473037, and U1631102. C. L. and L. D. are grateful for
support from the National Key Research and Development Program of
China through grant 2013CB834900 from the Chinese Ministry of Science
and Technology. This work has made use of data from the European Space
Agency (ESA) mission \textit{Gaia}
(\url{https://www.cosmos.esa.int/gaia}), processed by the
\textit{Gaia} Data Processing and Analysis Consortium (DPAC,
\url{https://www.cosmos.esa.int/web/gaia/dpac/consortium}). Funding
for the DPAC has been provided by national institutions, in particular
the institutions participating in the \textit{Gaia} Multilateral
Agreement. Also based on observations made with ESO Telescopes at the
La Silla Paranal Observatory under programme ID 380.D-0161.

\vspace{5mm} 
\software{PARSEC \citep[1.2S;][]{2012MNRAS.427..127B}, Astropy
  \citep{2013A&A...558A..33A}, Matplotlib \citep{2007CSE.....9...90H},
  SYNSPEC \citep{1992A&A...262..501H}}

\end{document}